\documentclass[a4paper,12pt,twoside]{cpc-hepnp}

\usepackage{bm}
\usepackage{graphics}
\usepackage{graphicx}
\usepackage{caption}
\usepackage{epsfig}
\usepackage{amssymb}
\usepackage{amsmath}
\usepackage{amssymb}
\usepackage{slashed}
\usepackage{multirow}
\usepackage{epsfig}
\usepackage[tight,hang]{subfigure}

\catcode`\ö=\active
\defö{\"{o}}

\catcode`\ý=\active
\defý{\i}

\newcommand{\be}{\begin{equation}}
	\newcommand{\ee}{\end{equation}}
\newcommand{\bea}{\begin{eqnarray}}
	\newcommand{\eea}{\end{eqnarray}}
\newcommand{\bdm}{\begin{displaymath}}
	\newcommand{\edm}{\end{displaymath}}
\newcommand{\baa}{\begin{array}}
	\newcommand{\eaa}{\end{array}}
\newcommand{\ds}{\displaystyle}

\newcommand{\qc}{\slashed{q}}

\newcommand{\ba}{\begin{eqnarray}}
\newcommand{\ea}{\end{eqnarray}}
\newcommand{\mM}{\mathcal{M}}
\newcommand{\mQ}{\mathcal{Q}}
\newcommand{\cu}{\mathpzc{u}}
\newcommand{\cv}{\mathpzc{v}}
\DeclareMathAlphabet{\mathpzc}{OT1}{pzc}{m}{it}

\begin{document}

\fancyhead[c]{\small arXiv : Preprint submitted and accepted in the journal Chinese Physics C} \fancyfoot[C]{\small \thepage}
		
		
		\title{The Role of the Neutrino Form Factors in the Energy Loss Rates of Pair Annihilation Process \thanks{The author would like to express my special thanks to Prof.Konstantin A.Kouzakov and Prof. A.Gutierrez-Rodriguez for the comments and useful discussions. }}
		
		\author{%
			C. Ayd{\i}n\email{coskun@ktu.edu.tr}
		}
		\maketitle
		
		\address{%
			Department of Physics, Karadeniz Technical University, 61080, Trabzon, Turkey
		}

		\begin{abstract}
			The stellar energy loss rates due to the production of the neutrino pair process $e^+e^- \rightarrow (W, Z, \gamma) \rightarrow \nu_e \overline{\nu_e}$  are calculated in the minimal extension  of Standard Model with electromagnetic properties of Dirac neutrinos which takes the account of the contribution of the neutrino charge radius, anapole moment and dipole moments. We have shown that the contribution of the electron neutrino's dipole moment is small compared with that of the charge radius. The obtained results are also considered with results obtained using the Standard Model.
		\end{abstract}
		
		
			
\section{Introduction}

Neutrinos are among the most fascinating and enigmatic particle in nature. They are also the second most abundant particles after photons in the visible universe. Different from the photons, the interaction of neutrinos are  extremely weakly. The neutrinos are the lightest of all known particles, they also don't interact with normal matter. When they are produced in stellar interior, they easily run away energy that would otherwise take much longer to be transported to the surface  by radiation or convection. The resulting energy sink in the center of the star can dictate the star's rate of nuclear burning, structure and evolution, and ultimately how its life ends. Any process which produces neutrinos in the stellar interior acts as a sink of stellar energy since the mean free path of neutrinos is much longer than the scale of stellar radius. The neutrinos are  valuable importance to high energy physics, astrophysics and cosmology \cite{ref1,ref2,ref3,ref4}. Gamov and Pontecorvo were the first to indicated the important role played by neutrinos in the evolution of stars. Therefore, the neutrino emissions process may affect the properties of matter at high temperature and also affect stellar evolution. The properties of neutrinos have become the subject of an increasing research afford for the sixty years. The possible electromagnetic properties of massive neutrinos include the charged radius, anapole moment, the dipole magnetic and electric moment  \cite{ref5,ref6,ref7,ref8,ref27n1,ref27n2,ref27_1,ref27_2}. The research for the neutrino mass, whether they are Dirac or Majorana particles, oscillations, and form factors especially the magnetic moment is of great significance for the choose of the theory of the elementary particles and for the clarifying of phenomena  such a supernova dynamics, stellar evaluation, and the production of the sun neutrino. There are many energy loss by neutrino pair production which is an important process in a wide range of astrophysical problems such as in the red giant stages of stellar evolution, neutron stars, supernova collapse, for cooling of white dwarfs \cite{ref9,ref10,ref11,ref12,ref13}.  While the three thermal neutrino processes which are plasmon decay ($\gamma_{plasmon} \rightarrow \nu_e + \overline{\nu}_e$), photo neutrino production ($e^-+\gamma \rightarrow e^- + \nu_e + \overline{\nu}_e$) and pair annihilation ($e^- + e^+ \rightarrow \nu_e + \overline{\nu}_e $) are most dominant, the bremstrahlung ($e^- + Z \rightarrow e^- + Z + \nu_e + \overline{\nu}_e$) and recombination processes ($e^-_{continuum} \rightarrow e^-_{bound} + \nu_e + \overline{\nu}_e$) play smaller role in astrophysics and cosmology. The energy loss rate of these process has been already calculated at various values of the temperature and matter density for different model \cite{ref14,ref15,ref16,ref17,ref18,ref19,ref20,ref21,ref22,ref23,ref24,ref25}. %

The most general expression for the effective vertex of the interaction $\nu\overline{\nu} \gamma$ is given as \cite{ref5,ref25,ref26,ref27} %
\be \label{eqn1}
\ds\Gamma_\mu = e \gamma_\mu F_1(q^2) \ds + (ie/2m_\nu)  \sigma_{\mu \lambda} q^\lambda F_2(q^2) +   (e/2m_\nu) \sigma_{\mu \lambda} q^\lambda \gamma_5  F_3(q^2)
+ e(\gamma_\mu -  \frac{{\qc}q_\mu}{q^2}) \gamma_5 q^2 F_4(q^2)
\ee
\noindent for the diagonal cases where $q^\lambda$ is the photon momentum and $\ds F_{1,2,3,4}(q^2)$ are the electromagnetic form factors of the neutrino. Note that in the derivations of the decomposition Eq.(\ref{eqn1}) the demands followed from the Lorentz invariance and electromagnetic gauge invariance are taken into account. The charge radius, magnetic moment, electric dipole moment and anapole moment are defined as \cite{ref8}
\be
\ds <r^2> = 6 \frac{d F_1(q^2) }{dq^2}|_{q^2=0},  \quad
{m_e \over m_\nu}F_2(q^2=0)=\mu_\nu, \quad {e \over 2m_\nu}F_3(q^2=0)=d_\nu, \quad F_4(q^2=0)=a.
\ee

A Dirac neutrino have the  charge, the magnetic moment, the electric dipole moment (which is absent in a CP invariant theory) and the anapole moment. However, a Majorana neutrino has only a diagonal anapole moment since its antiparticle is equal to itself(imposing the restriction of the C-,-P,-T-properties). Also, Majorana neutrinos can have as many transition moments as Dirac neutrinos.

 There have been studies on the energy loss of annihilation process in frame study of the SM and extension of the SM \cite{ref21,ref22,ref27_1,ref27_2}. In this study, we calculated the cross sections and energy loss rate of pair annihilation process in the extended SM includes neutrino electromagnetic form factors (especially  charge radius and anapole moment) effect on neutrino cross section takes in to account the $\gamma \nu \overline{\nu}$ vertex for the Dirac neutrinos \cite{ref7,ref8,ref27n1,ref27n2,ref25,ref26} which is not studied previously in literature as far as our knowledge in Section II. In Section III, we will obtain the numerical results and provide a discussion section.

\section{Calculation}

\begin{figure}[!h]
		\centering
		\subfigure[]{\includegraphics[height=4.0cm]{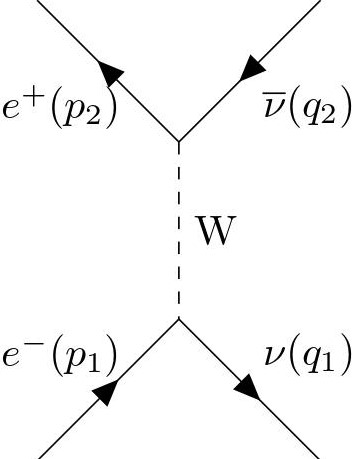}}
		\hspace{0.5cm}
		\subfigure[]{\includegraphics[height=4.0cm,width=5.0cm]{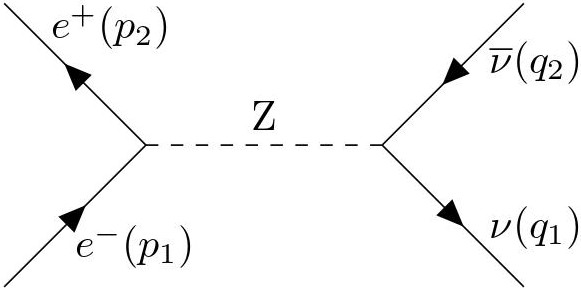}}
		\hspace{0.5cm}
		\subfigure[]{\includegraphics[height=4.0cm,width=5.0cm]{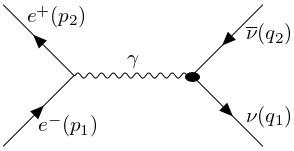}}
	\caption{The lowest order Feynman diagrams for the process $e^{+}+e^{-}\rightarrow \nu_e +\overline{\nu}_e$. The symbols in parantheses are the momenta of the particles. The black disc in (c) represents the interaction arising from the effective neutrino interaction beyond the SM.}
\label{fig1}
	\end{figure}

The lowest order diagrams for the process $e^-(p_1) + e^+(p_2) \rightarrow \nu_e(q_1) + \overline{\nu}_e(q_2) $  are given in Fig. \ref{fig1}. The matrix element of the process due to SM (Fig. \ref{fig1} (a)-(b)) in the low energy limit are given by making a Fierz transformation as

\be
\mM^{(SM)} = {i G_F \over \sqrt{2}}  \overline{\cu} (q_1) \gamma_\beta (1-\gamma_5) \cv(q_2) \overline{\cv}(p_2) \gamma^\beta (C_V-C_A \gamma_5) \cu(p_1)
\ee
where $C_V={1 \over 2} +2 \sin^2 \theta_W$ and $C_A={1 \over 2}$ for $\ds \nu_e$, and $C_V=-{1 \over 2} +2 \sin^2 \theta_W$ and $C_A=-{1 \over 2}$ for $\ds \nu_\mu$ and $\ds \nu_\tau$ because the electron neutrinos interact with both $W$ and $Z$ bosons but the muon and tau neutrinos interact only with $Z$ (see the case Fig. \ref{fig1} (b)).

The matrix element due to electromagnetic interaction (Fig. \ref{fig1} (c)) is given by \cite{ref28}
\be
\mM^{(\gamma)}= \mM^{(Q)} + \mM^{(\mu)}
\ee
with
\bdm
\mM^{(Q)} = {4\pi\alpha \over q^2} \overline{\cu}(q_1) (\gamma^\beta - {q^\beta {\qc} \over q^2}) \left[ {q^2 \over 6} <r_\nu^2> \right] \cv(q_2) \overline{\cv}(p_2) \gamma^\lambda \cu(p_1)
\edm
\bdm
\mM^{(\mu)} = -i {2\pi\alpha \over q^2} \overline{\cu}(q_1) \sigma^{\beta \lambda} q_\lambda  \left[ \mu \right] \cv(q_2) \overline{\cv}(p_2) \gamma^\lambda \cu(p_1)
\edm
where $\alpha=e^2/4\pi$, $\ds <r_\nu>=<r^2>+6 \gamma_5 a$ and $\ds \mu=\mu_\nu+i \gamma_5 d_\nu$. The effect of $\gamma_5$ is simply a multiplication by a factor of $-1$ in the above formulas for the ultra-relativistic neutrino.

The total matrix element of the process is the following
\be
\mM_t = \mM^{(SM)}+ \mM^{(Q)} + \mM^{(\mu)} .
\ee
There is no interference between the helicity-conserving ($\mM^{(SM)}$ and $\mM^{(Q)}$) and helicity-flipping ($\mM^{(\mu)}$) amplitudes. Combining the helicity-conserving amplitudes and using \mbox{$\ds q_\mu J^\mu(q)=0$}, we find
\be
\mM^{(SM)} + \mM^{(Q)}  = {i G_F \over \sqrt{2}}  \overline{\cu} (q_1) \gamma_\beta (1-\gamma_5) \cv(q_2)    \overline{\cv}(p_2) \gamma^\beta (C'_V-C_A \gamma_5) \cu(p_1)
\ee
where $C'_V=C_V + {\sqrt{2} \pi \alpha \over 3 G_F} <r_\nu^2>$.

When evaluating the cross section, even we have considered the neutrino mass, we ignored it in the calculations due to its negligible contribution. Then we obtain the following absolute total matrix element squared
\be
\baa{rcl}
|\mM_t|^2 & = & |\mM^{(SM)} + \mM^{(Q)}|^2 + |\mM^{(\mu)}|^2 \\\\
& = & \ds G_F^2 \left\{ {\alpha \over G_F^2} (\mu_\nu^2 + d_\nu^2) {(q_1 \cdot q_2)(q_2 \cdot p_1) \over (p_1+p_2)^2}  \right. \\ \\
& + & (C'_V + C_A)^2 (p_1 \cdot q_1)(p_2 \cdot q_2)\\ \\
& + & (C'_V - C_A)^2 (p_2 \cdot q_1)(p_1 \cdot q_2) \\ \\
& + & m_e^2 ({C'_V}^2 - {C_A}^2)(q_1 \cdot q_2) \left. \right\}
\eaa
\ee

Squaring the $\mM_t$, summing and averaging over the spins and the integrating over the final momenta by using Lenard's formula
\be
\ds \int {d^3 q_1 \over 2 w_1} {d^3 q_2 \over 2 w_2} \delta^4 (q-q_1-q_2) q^\mu q^\nu = {1 \over 24} \pi (2q^\mu q^\nu + g^{\mu \nu} q^2)
\ee
we have
\be
\baa{rcl}
\sigma v & =  & \ds {1 \over (2 \pi)^2}{1 \over 4 E_1 E_2} \int {d^3 q_1 \over 2 w_1} {d^3 q_2 \over 2 w_2} \delta^4 (p_1+p_2-q_1-q_2) \sum_s |\mM_t|^2 \\ \\
& = & \ds {G_F^2 \over 48 \pi E_1 E_2} \left\{ 8 {\alpha \pi\over G_F^2}(\mu_\nu^2 + d_\nu^2)(2m_e^2+p_1 \cdot p_2) + ({C'_V}^2+{C_A}^2)\left[2(p_1 \cdot p_2)^2 + 3m_e^2(p_1 \cdot p_2)+m_e^4\right] \right. \\ \\
& + & \ds \left. 12 ({C'_V}^2 -{C_A}^2) \left[ m_e^2 (p_1 \cdot p_2) + m_e^4 \right] \right\}
\eaa
\ee
where $E_1$ and $E_2$ are the energies of electron and positron, and $\overrightarrow{v}$ is the electron-positron relative velocity.

The rate of the energy loss per cubic centimeter per second $\mQ$ arising from the pair process depends not only on cross-section ($\sigma$) but also on the available densities of electrons and positrons. It is given by
\be
\label{eqn:el}
\ds \mQ =  {4 \over (2\pi)^6} \int {d^3 p_1 \over \textrm{exp}[(E_1-\mu_e)/k_B T]+1} {d^3 p_2 \over \textrm{exp}[(E_2+\mu_e)/k_B T]+1} (E_1 + E_2) v \sigma
\ee
where $\sigma$ is the process cross-section, $\ds {1 \over \textrm{exp}[(E_{1,2} \mp \mu_e)/k_B T]+1}$ is the Fermi-Dirac distribution functions for electron/positron, $\mu_e$ is the electron chemical potential, $T$ is the Stellar temperature and $k_B$ is the Boltzmann constant.We will assume that the star consists of a completely ionized gas in thermal equilibrium at a temperature $T$ with a density $\rho$. We will assume that the star consists of a completely ionized gas in thermal equilibrium at a temperature with a density

Now, according to \cite{ref13}, if we define $\lambda={k_B T \over m_e}$ and $\nu={\mu_e \over k_B T}$ such as $m_e$ is the electron mass, the we will get the functions
\be
\label{eqn:int}
\ds G_n^\pm(\lambda,\nu) = \lambda^{3+2n} \int_{\lambda^{-1}}^ \infty dx {x^{2n+1}(x^2-\lambda^{-2}) \over \textrm{exp}(x\pm\nu)+1 } .
\ee

The energy loss rate for all values of $\lambda$ and $\nu$ in Eq. \ref{eqn:el} can be written in terms of the integrals given Eq. \ref{eqn:int} as
\be
\baa{rcl}
\mQ & = & \ds {G_F^2 m_e^9 \over 18 \pi^5} \left\{ {12 \pi \alpha \over G_F^2 m_e^2} (\mu_\nu^2+ d_\nu^2) \left[2 (G_0^+G_{-1/2}^- + G_0^-G_{-1/2}^+) + G_0^-G_{1/2}^+ + G_0^+G_{1/2}^- \right] \right. \\ \\
& + & 2 ({C'_V}^2 + {C_A}^2) \left[5 (G_0^+G_{-1/2}^- + G_0^-G_{-1/2}^+) + 7 (G_0^-G_{1/2}^+ + G_0^+G_{1/2}^-) \right. \\ \\
& - & \left.  (G_1^-G_{-1/2}^+ + G_1^+G_{-1/2}^-) + 8 (G_1^-G_{1/2}^+ + G_1^+G_{1/2}^-) \right] \\ \\
& + & \left.  36 ({C'_V}^2 - {C_A}^2) \left[ G_0^+G_{-1/2}^- + G_0^-G_{-1/2}^+ + G_0^-G_{1/2}^+ + G_0^+G_{1/2}^- \right]   \right\} .
\eaa
\ee

Unfortunately, these integrals can not be calculated analytically for any values of $\lambda$ and $\nu$.We  have evaluated them for special ranges of temperatures and densities. In order to compare the results previously  obtained in the SM for electron-positron pair annihilation we  evaluate $\mQ$ in various regions of $\lambda$ and $\nu$, and also calculate the ratio ${\delta \mQ \over \mQ^{SM}} = { \mQ - \mQ^{SM} \over \mQ^{SM}}$.

In region I ($\rho \le 10^5 gr/cm^3, 3 \times 10^8 \, ^oK \le T \le 3 \times 10^9 \, ^oK$); $\lambda  \ll 1, \nu \ll 1/\lambda$ (nonrelativistic and nondegenerate case)
\be
\mQ_1 = {G_F^2 \over \pi^4} m_e^6 T^3 e^{-2m_e/T} \left[ { \alpha  \pi  \over G_F^2 m_e^2} (\mu_\nu^2 + d_\nu^2) + {C'}_V^2 \right]
\ee
 and
in region II ($ 10^4 gr/cm^3 \le \rho \le 10^6 gr/cm^3,  T \le 3 \times 10^8 \, ^oK$); $\lambda  \ll 1,  1/\lambda < \nu < 2/\lambda$ (nonrelativistic and mildly degenerate case)
\be
\mQ_{II} = {\sqrt{2\pi}G_F^2 \over  \pi^3} m_e^6 \left( {\rho \over \mu_e} N_A  \right) \left( {T \over m_e} \right)^{3/2}   e^{-(m_e+\mu_e)/T} \left[ { \alpha  \pi  \over G_F^2 m_e^2} (\mu_\nu^2 + d_\nu^2) + {C'}_V^2 \right]
\ee
then we get
\bdm
{\delta \mQ_{I} \over \mQ^{SM}_{I}} = {\delta \mQ_{II} \over \mQ^{SM}_{II}} = { [C_V + \ds {\sqrt{2}\pi \alpha \over 3 G_F} < r_\nu^2> ]^2 + \ds { \pi \alpha \over G_F^2 m_e^2} (\mu_\nu^2 + d_\nu^2 )  \over C_V^2 } -1 .
\edm

Similarly, in region III ($ 10^7 gr/cm^3 \le \rho , \,  6 \times 10^7 \, ^oK < T$); $\lambda  \ll 1,  1 \ll \lambda \nu$ (relativistic and nondegenerate case),
\be
\mQ_{III} = {\sqrt{2\pi}G_F^2 \over 20 \pi^3} m_e^4 \mu_e^2 \left( {\rho \over \mu_e} N_A  \right) \left( {T \over m_e} \right)^{3/2}   e^{-(m_e+\mu_e)/T} \left[ {C'}_V^2 + C_A^2\right]
\ee
 in region IV ($ 10^7 gr/cm^3 \le \rho$); $1 \ll \lambda ,  \nu \ll 1$ (relativistic and nondegenerate case)
\be
\mQ_{IV} = {7 \zeta(5) G_F^2 \over 12 \pi} (T)^9 \left[ {C'}_V^2 + C_A^2\right]
\ee
  and in region V ($ 10^8 gr/cm^3 \le \rho$, $10^{10} \, ^oK \le T$  (at the lowest), $ 10^{10} gr/cm^3 \le \rho$, $10^{10} \, ^oK \le T \le 10^{11} \, ^oK$ (at the extendable)); $1 \ll \lambda ,  1 \ll \nu$ (relativistic and degenerate case)
\be
\mQ_V = {2G_F^2 \over 5 \pi^3} (T)^4 \mu_e^2 \left( {\rho \over \mu_e} N_A  \right)    e^{-\mu_e/T} \left[ {C'}_V^2 + C_A^2\right]
\ee
and again one can get
\bdm
{\delta \mQ_{III} \over \mQ^{SM}_{III}} = {\delta \mQ_{IV} \over \mQ^{SM}_{IV}} = {\delta \mQ_{V} \over \mQ^{SM}_{V}} = { [C_V + \ds {\sqrt{2}\pi \alpha \over 3 G_F} < r_\nu^2> ]^2 + C_A ^2   \over C_V^2 + C_A^2 } -1 .
\edm
 Noticed that, the typical approximation for these regions ($III-V$) only considers the terms of dominant powers, so there is no dependence on the $\mu_\nu$ and $d_\nu$ of the neutrino.

\section{Numerical Results and Discussion}

Since the magnetic moment contribution has been already studied by many researchers (\cite{ref21,ref22} and references there in), as originality, we will try to obtain the effects of form factors numerically from the proposed analytical expressions especially charge radius (or anapole moment) of the Dirac neutrino on the energy loss rates through the physical process of pair-annihilation $e^+ + e^- \rightarrow \nu_e + \overline{\nu}_e $. 
 
As is known,  this process is one of the main mechanisms of neutrino pair production relevant for the neutrino luminosity. We obtained the approximated formula for energy loss($Q$) and the correction contribution in comparing with that of the SM. We investigate for both degenerate and non-degenerate Fermi gas for the case $d_\nu=0$ using the values of the parameters given in \cite{ref21,ref22,ref29,ref30,ref31,ref32,ref33}.

\begin{figure}[!h]
		\centering
		\includegraphics[height=8cm]{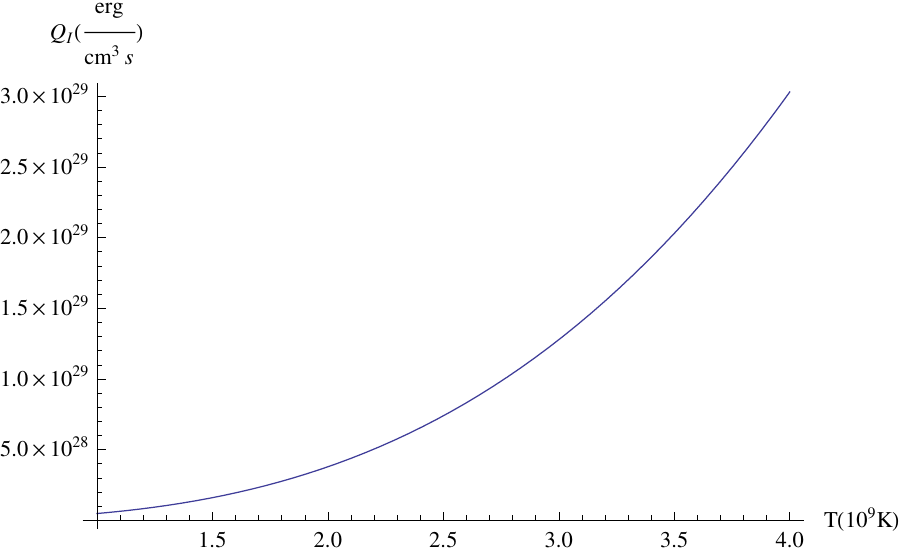}
	\caption{The energy loss rate($Q_I$) as a function of temperature($T$) for the case $\mu_\nu=1.58 \times 10^{-12} \mu_B$ and $<r_\nu^2>=1.5 \times 10^{-32} cm^2$}
\label{fig_02}
\end{figure}

The dependence of $Q_I$ as a function of temperature ($T$) is visualized in Fig. \ref{fig_02}. It is seen that as the temperature increases the stellar energy loss rate also increases. 

\begin{figure}[!h]
		\centering
	\subfigure[$\mu_\nu = 0$]{\includegraphics[width=8.5cm]{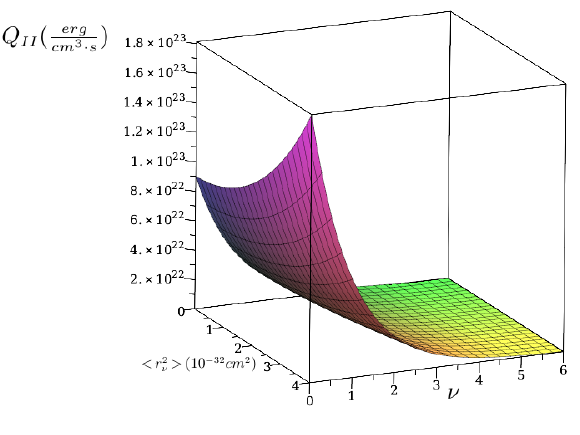}}
	\subfigure[$<r_\nu^2> = 0$]{\includegraphics[width=8.5cm]{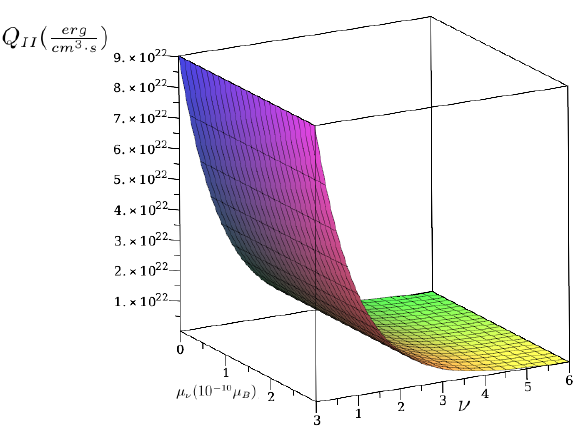}}
	\caption{The energy loss rate($Q_{II}$) as a function of $\ds \nu={\mu_e \over k_B T}$  and charge radius ($<r_\nu^2>$), and $\ds \nu$ and magnetic moment ($\mu_\nu$) for the case $T=3 \times 10^8 K$ and $\rho= 1.1 \times 10^6 gr/cm^3$}
\label{fig_03}
\end{figure}

In Fig \ref{fig_03} the dependence of $Q_{II}$ on the pair of parameters of charge radius with the degeneration parameter $\nu$ (Fig \ref{fig_03}-(a) in this case the magnetic moment is taken as zero) and magnetic moment with the degeneration parameter $\nu$ (Fig \ref{fig_03}-(b) in this case the charge radius is taken as zero) are displayed. As it is seen that $Q$ decreases with the increase of $\nu$ since $\nu$ is inverse proportional to the temperature $T$. Also the effect of the charge radius on the energy loss compared to the magnetic moment is observed more clearly.

\begin{figure}[!h]
	\centering
	\subfigure[]{\includegraphics[width=8.5cm]{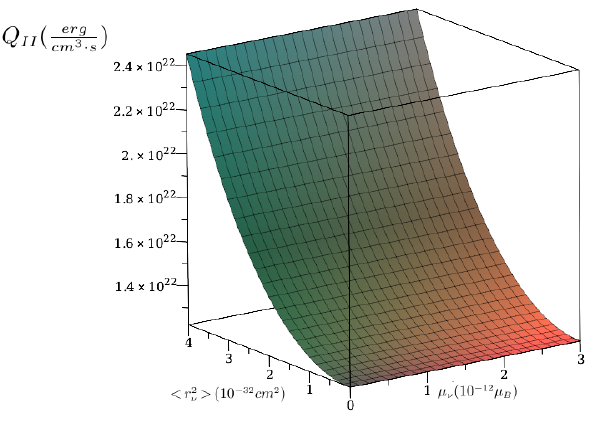}}
	\subfigure[]{\includegraphics[width=8.5cm]{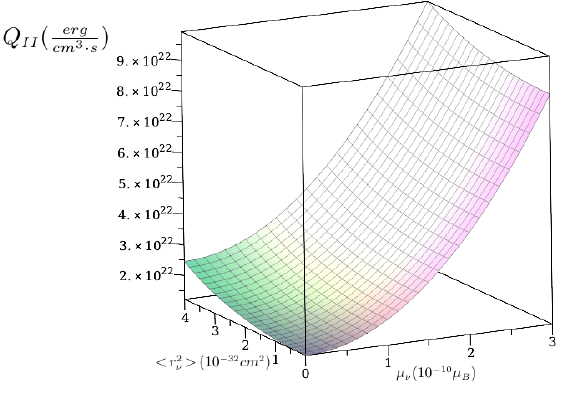}}
	\caption{The energy loss rate($Q_{II}$) as a function of magnetic moment ($\mu_\nu$) and charge radius ($<r_\nu^2>$) for the case $T=3 \times 10^8 K$ and $\rho= 1.1 \times 10^6 gr/cm^3$}
	\label{fig_045}
\end{figure}

We have compared the effect of the change in order of the magnetic moment from $10^{-12} \mu_B$ to $10^{-10} \mu_B$ on $Q_{II}$  in Fig \ref{fig_045}. It is seen that as the magnetic moment is getting large even the order of the energy loss rate stays almost same level, again the effect of the magnetic moment  is more pronounced for the large value case (Fig \ref{fig_045}- (b)) for the dominant of the contribution obtained from the standard model. In the case of the magnetic moment of the electron neutrino, the best bound is derived from globular cluster red giants energy loss  $\mu_{\mu_e} < 3 \times 10^{-12} \mu_B$ \cite{ref21,ref30,ref34}. If one takes magnetic moment with the order $10^{-14} \mu_B$ \cite{ref39}, the effect of the considered formulation  in the numerical calculation will not observed effectively. The contribution of the charge radius still exist in the calculations.

\begin{figure}[!h]
		\centering
		\includegraphics[height=8cm]{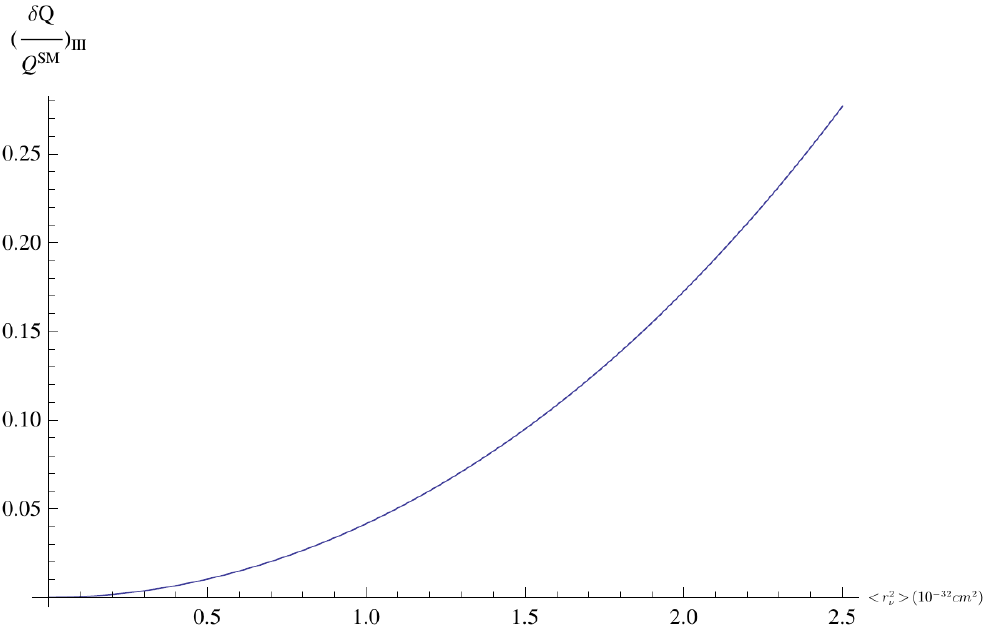}
	\caption{The ratio $({\delta \mQ \over \mQ^{SM}})_{III} = ({ \mQ - \mQ^{SM} \over \mQ^{SM}})_{III}$ as a function of charge radius ($<r_\nu^2>$)}
\label{fig_06}
\end{figure}

Finally, in Fig \ref{fig_06} we have displayed the behavior of ${\delta \mQ \over \mQ^{SM}}$ with respect to charge radius for the cases $III, IV$ and $V$. It is already stated that ratio depends only to charge radius due to the negligible effect of dipole moment ($\mu_\nu$).

As a conclusion, in this study we have calculated the charge radius and magnetic moment effects numerically for the electron neutrino. If $\tau$ neutrinos are considered, where their magnetic moments are $10^6$ times grater then electron neutrino,  the effect of the indicated term become dominant and the contribution of the magnetic moment becomes bigger than the contribution of the charge radius for this situation. As a further study for the extension, the similar calculations can be done for the plasmon decay and photo neutrino production.

\clearpage

\end{document}